\documentclass[12pt,preprint]{aastex}

\newcommand{\chandra}{{\it Chandra}}

\def\sgr{SGR~1627$-$41}
\newcommand{\etal}{{\it et al.~}}

\begin{document}

\title{Unraveling the cooling trend of the Soft Gamma Repeater, SGR~$1627-41$}

\author{C. Kouveliotou\altaffilmark{1,2}, D. Eichler\altaffilmark{3}, P. M. Woods\altaffilmark{2}, Y. Lyubarsky\altaffilmark{3}, S. K. Patel\altaffilmark{4}, E. G\"o\u{g}\"u\c{s}\altaffilmark{2,5}, M. van der Klis\altaffilmark{6}, A. Tennant\altaffilmark{1}, S. Wachter\altaffilmark{7}, K. Hurley\altaffilmark{8}}

\email{chryssa.kouveliotou@nasa.gov}

\altaffiltext{1} {NASA/Marshall Space Flight Center, NSSTC, SD-50, 320
Sparkman Drive, Huntsville, AL 35805, USA}
\altaffiltext{2} {Universities Space Research Association, NSSTC,
SD-50, 320 Sparkman Drive, Huntsville, AL 35805, USA}
\altaffiltext{3} {Ben Gurion University, Physics Department, POB 653, Beer Sheva 84105, Israel}
\altaffiltext{4} {National Research Council Fellow, NSSTC, SD-50, 320
Sparkman Drive, Huntsville, AL 35805, USA}
\altaffiltext{5} {Sabanci University, FENS, Orhanli-Tuzla, Istanbul 34956, Turkey}
\altaffiltext{6} {Astronomical Institute ``Anton Pannekoek''and Center
for High Energy Astrophysics, University of Amsterdam, Kruislaan 403,
1098 SJ Amsterdam, The Netherlands}
\altaffiltext{7} {SIRTF Science Center, Caltech M/S 220-6, 1200 E. California Blvd., Pasadena CA 91125, USA}
\altaffiltext{8} {University of California, Berkeley, Space Sciences Laboratory, Berkeley, CA 94720-7450, USA}

\begin{abstract}

\sgr\ was discovered in 1998 after a single active episode which lasted $\sim6$ weeks. We report here our monitoring results of the decay trend of the persistent X-ray luminosity of the source during the last 5 years. We find an initial temporal power law decay with index 0.47, reaching a plateau which is followed by a sharp (factor of ten) flux decline $\sim800$ days after the source activation. The source spectrum is best described during the entire period by a single power law with high absorption ($N_{\rm H}=9.0(7)\times10^{22}$ cm$^{-2}$); the spectral index, however, varies dramatically between $2.2-3.8$ spanning the entire range for all known SGR sources. We discuss the cooling behavior of the neutron star assuming a deep crustal heating initiated by the burst activity of the source during 1998.  

\end{abstract}


\slugcomment{Accepted - ApJ Letters}

\section{Introduction}

Soft gamma repeaters (SGRs) are a rare subclass of neutron stars
characterized by their emission of randomly recurring outbursts of
hard X- and soft $\gamma$-rays. There are currently four SGRs
identified in our galaxy and one in the Large Magellanic Cloud
(LMC), three of which have been found to pulse with periods
ranging between $5-8$ s (for a review see Kouveliotou 2004). The very rapid secular increase of these periods, (spin down of $\dot{P}\sim10^{-10}$ s/s), argues for
angular momentum loss from a highly magnetized neutron star
($B\sim10^{14}$ G). This idea was developed by
Duncan and Thompson (1992, henceforth DT92) and,
subsequently, by Paczynski (1992), following the detection of the
most intense high energy transient observed to date, the giant
flare of 1979 March 5 (from SGR~$0526-66$ in the LMC), which gave
a lower limit of the source magnetic field of $\sim10^{14}$ G.
DT92 dubbed such sources ``magnetars''.

\sgr\ was discovered with the Burst And Transient Source Experiment (BATSE) on
the Compton Gamma-Ray Observatory (CGRO) in June 1998 (Kouveliotou \etal 1998), when it
emitted over 100 bursts within an interval of 6 weeks (Woods \etal 1999); no further
burst emission has been observed to date (July 2003). Roughly 20\% of all events,
representing $\sim98\%$ of the total burst-emitted energy (20 -150 keV), were
bunched in an interval of three days, 1998 June $15-18$. The X-ray counterpart
to \sgr, SAX~J$1635.8-4736$, was discovered in a BeppoSAX/Narrow Field Instrument
(NFI) observation
on 1998 August 7, at $\alpha = 16^{\rm h} 35^{\rm m} 49.8^{\rm
s}$ and $\delta = -47^{\circ} 35^{\prime} 44^{\prime\prime}$ (J2000) with an
error circle of radius 1$^{\prime}$ (95\% confidence level; Woods \etal 1999). In Wachter \etal (2003) we report an improved (error radius $\lesssim0.3^{\prime\prime}$) source
location recently derived with the \chandra\ X-Ray Observatory, together
with results of our near-infrared searches for a counterpart.  A search for coherent
pulsations in the BeppoSAX dataset, when the source intensity was higher,
showed marginal evidence ($\sim3\sigma$ confidence level) near 6.4 s; to
date the source spin period remains unknown.

During their quiescent periods, SGRs have been identified as persistent X-ray
sources with luminosities of $\sim10^{34-35}$ ergs/cm$^2$s. When active, their
outbursts last anywhere from days to a year, with different burst frequency and
intensity per source and per outburst. SGR bursts have typical durations
of 0.1 s and their spectra are usually best-fitted to an optically thin thermal
bremsstrahlung function with $kT\sim30$ keV. The burst size distributions
follow a power law of index $-0.6$, with peak luminosities of events ranging
from the small, common bursts at $\lesssim$5$\times10^{37}$ erg/s to the
rare, giant flares (only two have been observed) at $\sim10^{44}$erg/s.

The enhancing effect of the SGR bursting activity on the flux
level of their persistent emission is well documented for
SGR~$1900+14$ where the flux was shown to increase by a factor of $\sim$700 after the giant flare of 1998 August 27. The flux from
the source decayed within $\sim40$ days according to a
power-law, $F \propto (t-t_0)^{-0.7}$ (Woods \etal 2001).  For the next several
months, the burst activity and persistent/pulsed flux level
gradually declined.  The SGR was not observed to reach its
quiescent flux level of $1\times10^{-11}$ erg/cm$^2$s until 2000 April, approximately two years after the burst reactivation of the
SGR, though extrapolation of the 40-day decay after 1998 August 27
suggests that it may have done so between observations. The
source has been intermittently active ever since,
resulting in erratic flux decay behaviour.

In contrast, \sgr\ has remained dormant after a single active
episode thus being a better long term target for SGR cooling
studies. We present here (Sections 2, 3, 4, and 5) the results of
our monitoring of the flux decay of the quiescent X-ray
counterpart of \sgr\ obtained with imaging instruments (BeppoSAX,
ASCA and \chandra) spanning an interval of roughly five years. In
Section 6 we discuss the source's atypical decay within the
framework of neutron star crust cooling and its implications on
the properties of SGR progenitors.

\section{BeppoSAX Observations}

Since 1998, we have observed the source four times with the BeppoSAX/NFI.
Results from the first two observations were presented in Woods \etal (1999).
Here, we present results from a refined analysis of the earlier data together
with the two subsequent observations in August 1999 and September 2000. For all
BeppoSAX data sets, we followed the data extraction procedure described in
Woods \etal (1999). A log of the observations and spectral fit results is
given in Table 2. Due to the significant contribution to the background from
the Galactic ridge, we are using contemporaneous data from a concentric
annulus  around our source position for our background spectrum.  An energy
dependent multiplicative factor has been applied to the  background  spectrum
to correct for reduced efficiency at off-axis angles in both the  LECS and
MECS. From these earlier BeppoSAX observations, where the higher
source intensity allowed for better statistics, we have determined using XSPEC (v11.2; Arnaud 1996) that, in all cases, a single power-law (PL) model is the best fit spectral function for the data.

\section{ASCA Observations}

\sgr\ was observed by the Advanced Satellite for Cosmology \& Astrophysics ({\it
ASCA}) on 1999 February 26-28.  Onboard {\it ASCA} are four independent X-ray
telescopes and four independent X-ray detectors; the latter are two Gas Imaging
Spectrometers (GIS) and two Solid-State Imaging Spectrometers (SIS).  For the
present analysis, we have extracted all available GIS and SIS data sets and
applied the standard screening criteria\footnotemark{}\footnotetext{http://heasarc.gsfc.nasa.gov/docs/asca/screening.html}.
We reduced the data
separately for each GIS and for each SIS detector in 1-CCD mode, and
extracted spectra using XSELECT (v2.1) from circular regions centered on the
SGR with radii of $4\arcmin$ and $2.6\arcmin$ for GIS and SIS, respectively.
Since the source was faint, we used a smaller region than the recommended {\it
ASCA} source extraction region to reduce the background. Finally, the background
was taken from neighboring regions on each detector.

To derive an accurate measure of the source flux, we generated off axis GIS
ancillary response files, dead-time corrected the GIS data, and included
only the high and medium bit rate SIS data.  We then generated response files
(using FTOOLS v5.2 tasks {\it ascaarf} and {\it sisrmg}) for each detector
separately.  We combined all the SIS data into a single data set, (similarly
the GIS data), weighting in each case the response functions by their
individual effective live times. We grouped the data into energy
channels that contained at least 25 events each.

We performed independent fits to the SIS and GIS data (using XSPEC) of a PL model including the effects of Galactic absorption.  We note that the final flux we measure for the SGR (see also section 5 and Table 2) is significantly lower than the value reported earlier ($F=5.1\times10^{-12}$ ergs/cm$^2$ s) by Hurley \etal (2000). This discrepancy is due to the low source count rate in the GIS detectors, which results in a critical dependence of the measured flux on the
selected background region.  For example, choosing a background region several
arcminutes away from the SGR, where the detector count rate is much lower,
we find a flux comparable to the value reported by Hurley \etal (2000).
In contrast, we have selected a background region close in angle to the SGR, which resides in an
area of the sky that contains significant diffuse emission.  This choice is justified by the fact that the flux we measure with the GIS is in
good agreement with the flux measured by the SIS detectors (within 10\%), which
do not suffer from this issue due to their smaller field-of-view.

\section{\chandra\ observations}

We observed \sgr\ with \chandra\ on 1999 September 20 (Obs1) and on 2003 March 24
(Obs2). In both observations the SGR fell on ACIS-S3 (operating in Timed
Exposure mode), a back-illuminated CCD with good spectral resolution. We used
calibration products from \chandra\ CALDB (v2.21). Using the standard CIAO
(v2.3) tools, we applied the appropriate gains, CTI corrected the events, and
filtered the data to include events with grades=0,2-4,6. We selected a circular
extraction region of $3\arcsec$ radius centered on the SGR for the source
spectra.  We collected the background spectrum from annular regions centered on
the SGR with inner and outer radius of $3\arcsec$ and $30\arcsec$,
respectively. The $2-10$ keV background-subtracted source count rates were
0.0049 cnt/s (Obs1) and 0.0058 cnt/s (Obs2) (the background contributed only
$\sim3\%$ to the total count rate for each observation). We then generated the
response files using standard CIAO tools and corrected the ancillary response
files to account for the time dependent degredation of the ACIS quantum
efficiency. We have investigated
bremsstrahlung, blackbody, and PL models using XSPEC and find that the PL model gives the least
C-statistic for both data sets (fitted individually or
simultaneously). We have, therefore, adopted the absorbed PL model as the
preferred continuum model for the \chandra\ observations.
Our spectral fit results to the ungrouped data using the C statistic are
given in Table 1. A simple ratio of the two data sets indicates that Obs2 has a softer spectrum, as also indicated by the spectra in Table 1.

\begin{table}[!h]
%\begin{minipage}{1.0\textwidth}
\begin{center}
\caption{\chandra/ACIS Power-Law Model Fit Results}
%\vspace{8pt}
\begin{tabular}{cccccc} \hline \hline

Parameter & OBS1 & OBS2 & OBS1$+$OBS2$^{d}$ \\\hline

$N_{\rm H}(10^{22}$ atoms cm$^{-2})$ & 9(2)$^{a}$ & 10(2) & 9.6(1.1)\\

Index  & 2.2(5) & 3.3(6) &  2.3(4), 3.1(4) \\
Flux$^{b}$ & 0.33 & 0.44 & 0.34, 0.39 \\
C-Stat (Goodness$^{c}$) & 447.9 (86.6\%) & 369.1
(92.6\%) & 817.3 (95.2\%) \\

\hline\hline
\end{tabular}
\end{center}

\noindent$^{a}$ Uncertainties are given at the 68\% confidence level for 1 parameter $^{b}$ Unabsorbed 2-10 keV flux in units of $10^{-12}$ ergs/cm$^2$s 

\noindent $^{c}$ Percentage of 1000 realizations of the model in which C-stat is less then that listed in table $^{d}$ Obs1 and Obs2 are simultaneously fitted with $N_{\rm H}$ forced to have the same value for both data sets and the remaining parameters are kept free

%\end{minipage}\hfill
\end{table}

\section{Joint Spectral Analysis--Source energetics}

In Sections 2-4, each of the seven data sets was fit independently to a PL model
attenuated by interstellar absorption. Motivated by the consistency of
the measured spectral parameters from all seven observations, we performed a
simultaneous fit to all data sets linking the Hydrogen column density, $N_{\rm H}$. We have kept the value of $N_{\rm H}$ linked assuming that there is no obvious
physical reason for the intrinsic source absorption to change while the source
is in quiescence. For all observations, the fit was
statistically acceptable ($\chi^2/\nu=690/655$). We measure the (linked) effective hydrogen column to be $N_{\rm H} =
(9.0\pm0.7)\times10^{22}$ cm$^{-2}$. Table 2 lists all other fit parameters.

\begin{table}[!h]
%\begin{minipage}{1.0\textwidth}
\begin{center}
\caption{Observation log and linked spectral fit parameters for \sgr.}
%\vspace{10pt}
\begin{tabular}{cccccc} \hline \hline

Mission   & Date & Exposure &  Photon   & Flux$^{a}$  \\
 & (MJD)  & (ksec) &  Index & (10$^{-12}$ ergs/cm$^2$s$^1$) \\\hline

 BeppoSAX & 51032.25 &  44.9 &   2.54(16) & 6.56 \\
 BeppoSAX & 51072.52 &  30.4  &   2.77(21) & 4.94 \\
 ASCA     & 51235.20 &  67.9 &   3.24(24) & 2.76 \\
 BeppoSAX & 51399.78 & 80.4 &   3.77(30) & 2.29 \\
 BeppoSAX & 51793.59 & 61.3 &   2.99(29) & 1.86 \\
 Chandra  & 52182.50 & 48.9 &   2.17(30) & 0.267 \\
 Chandra  & 52505.00 & 25.7 &   2.95(36) & 0.266 \\

\hline\hline
\end{tabular}
\end{center}
\noindent$^{a}$ Unabsorbed $2-10$ keV flux
%\end{minipage}\hfill
\end{table}

Figure 1 displays the evolution of the source flux (upper panel) and spectral index (lower panel) since its activation. During the first $\sim$800 days following the 1998 outburst, the flux from \sgr\
decayed monotonically as a power-law with exponent $\alpha=0.47$. The last
BeppoSAX observation indicated that the source might have reached a plateau,
approximately two years after activation. However, the two subsequent \chandra\ observations (at $\sim$1200 and $\sim$1500 days after the outburst) are significantly below (almost a factor of ten) the extrapolated decay trend. The photon index does show significant changes in the $\sim5$ years following the 1998 outburst; the probability that the observed variability is due to statistical fluctuations is
7.1 $\times$ 10$^{-4}$.  We discuss these results further in the next section.

While the spectra of SGRs $1900+14$ and $1806-20$ are relatively
hard, with indices between 2.0 and 2.5, SGR~$0526-66$ is much
softer with a spectral index of 3.5. Interestingly, the spectral
index range of \sgr\ ($2.2-3.8$) spans all other SGR sources
observed at their quiescent states, although we do not see the
blackbody (bb) component observed in SGR~$1900+14$. This is
not unexpected, given that the bb contribution in the unabsorbed flux of 1900$+$14 is of
the order of 20\%; since the column density along the line of
sight for \sgr\ is almost 10 times higher, extreme absorption
between 0.1 - 2.0 keV would hinder the detection of any bb
component that contributes less than at least 50\% of
the (unabsorbed) flux.

Near the source activation the flux was $6.56\times10^{-12}$
ergs/cm$^2$s, corresponding to a source luminosity of
$L=9.5\times10^{34}$ ergs/s (assuming a source distance of 11
kpc; Corbel \etal 1999). The luminosity decayed slowly to $2.7\times 10^{34}$ before
it plummeted to its current value of $3.9\times 10^{33}$ erg/s. In
comparison, the persistent luminosity levels of SGRs $1900+14$,
$1806-20$ and $0526-66$ are $\sim6\times10^{34}$, 4.1$
\times10^{35}$ and $\sim10^{36}$ ergs/s, respectively. If \sgr\
decays further below its current luminosity level, it may well
provide the first direct link between SGRs and Isolated Neutron
Stars, whose luminosities it seemed to be approaching pretty
rapidly through day 1200 (Figure 1).

\begin{figure}
\figurenum{1}
\epsscale{1.0}
%\plotone{spec_hist2_1627_fit.ps}
\plotone{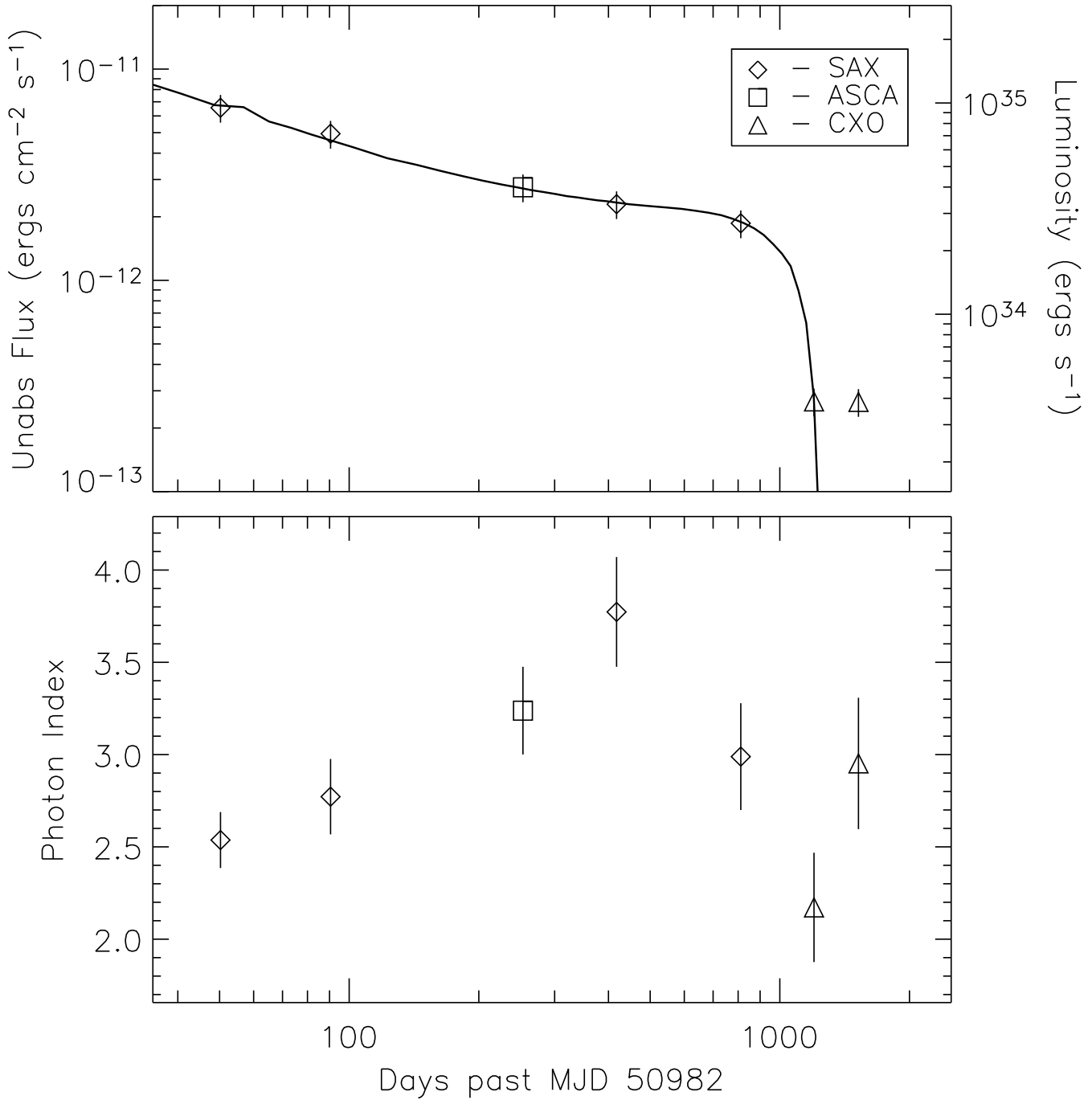}
\vskip0.1truein

\caption{{\it upper panel:} Flux history of \sgr\ derived with three different
spacecraft over $\sim$1500 days ($2-10$ keV). Errors in the flux are assumed to be 15\%. The solid line is the numerical fit to the data as discussed in Section 6. {\it lower panel:} Evolution of
the power law spectral index of the source during the same interval.}

\end{figure}

\section{Interpretation and Discussion}

The study of the afterglow following an SGR episodic energy release is complicated by the
fact that both the magnetosphere and the surface are subject to change, and by
the
unpredictable nature of reheating by additional, later bursts. Conventional
wisdom,
though still somewhat tentative,
would suggest that the surface radiation is reprocessed and non-thermalized by
resonant scattering in the magnetosphere (Thompson, Lyutikov and Kulkarni 2002)
This suggests that sharp spectral changes would signify changes in the
 magnetospheric configuration; conversely, an ordered, gradual
decline in intensity
with a constant spectrum indicates a decline in the thermal surface emission.
To study the latter, a period of gradual decline is needed uninterupted by new
events that reheat the
 surface. It was not obvious {\it a priori} that all these factors could be
unraveled.

The 40-day afterglow following the 1998 August 27 giant flare of
SGR~$1900+14$
was extremely well fit by a crustal cooling model (Lyubarsky, Eichler and
Thompson 2002, hereafter LET),  despite the existence of many small
bursts that took place during that period. Ironically, the
luminosity of this source was seen to increase and behave
erratically over the following  year, when there was relatively
little SGR burst activity. In other words, SGR bursts do not
necessarily heat the crust significantly and, moreover, there may be other kinds of sporadic heating  that is not expressed in SGR bursts.

Encouraged by the success of our outer crust cooling model for the 40-day afterglow
of SGR~$1900+14$, we
attempt here to understand the three year monotonic decline of \sgr\ as cooling after a single
deep crustal heating event coinciding with the burst activity of 1998.
Details of our calculational methods can be found in LET. In particular,
the leveling of the flux during the third year followed by its sharp
decline are curious features that beg for an explanation within this model.

We present in Figure 2 our numerical calculations of the temperature evolution (cooling)
of the neutron star crust with depth, $z$, assuming an initial energy injection to the crust of the order of 10$^{44}$ ergs (estimates of the total energy released in bursts during
the activation of \sgr\ range between $4\times10^{42}-2\times10^{43}$ ergs; here we assume that the conversion efficiency of the total energy released during the activation into soft gamma-rays is considerably less than 100\%). The general shape of the initial temperature profile was chosen under the assumption that the energy density
of deposited heat varies with height more slowly than the specific heat: thus,
in the outer crust, where the specific heat increases with depth,  the initial
temperature (i.e. immediately after the  heating event) declines with depth.  In the
inner crust, on the other hand, the specific heat rises even more rapidly with the onset
of neutron drip, but only if these neutrons
are unpaired. (Paired neutrons do not represent new degrees of freedom because
 they are condensed into a superfluid.)
  Below $T_c$, the  temperature below which neutrons pair,
  the specific heat of the crust material is small, and it
rises dramatically near $T_c$ as free, unpaired  neutrons appear.
It is thus reasonable to choose the initial temperature to be near
$T_c$ in the lower crust. The theoretically predicted behavior of
$T_c$ as a function of density can vary from one model to the
next, but the qualitative feature is that it increases with
density and then decreases as the nature of the pairing changes, hence we put a bump in the initial temperature profile at $400<z<500$ m. The exact shape of the bump is not very important as it gets washed out before it affects the surface temperature.

We also assume that the core  temperature is very low. This is justified if the mass of the neutron star exceeds $1.5M_{\odot}$ and the density, therefore, is high enough at the core to
cool it via the direct URCA process. The assumption of a high mass neutron
star makes the additional prediction of a thin crust, quite independently of the
low core temperature, and this is reflected in the relatively rapid
timescale (a few years) for cooling of the inner crust.  If our
interpretation of the transient cooling  is correct,  we thus determine the
mass of \sgr\ via two independent considerations to be above $1.5M_{\odot}$.

\begin{figure}
\figurenum{2}
\epsscale{0.8}
%\plotone{temp43.ps}
\plotone{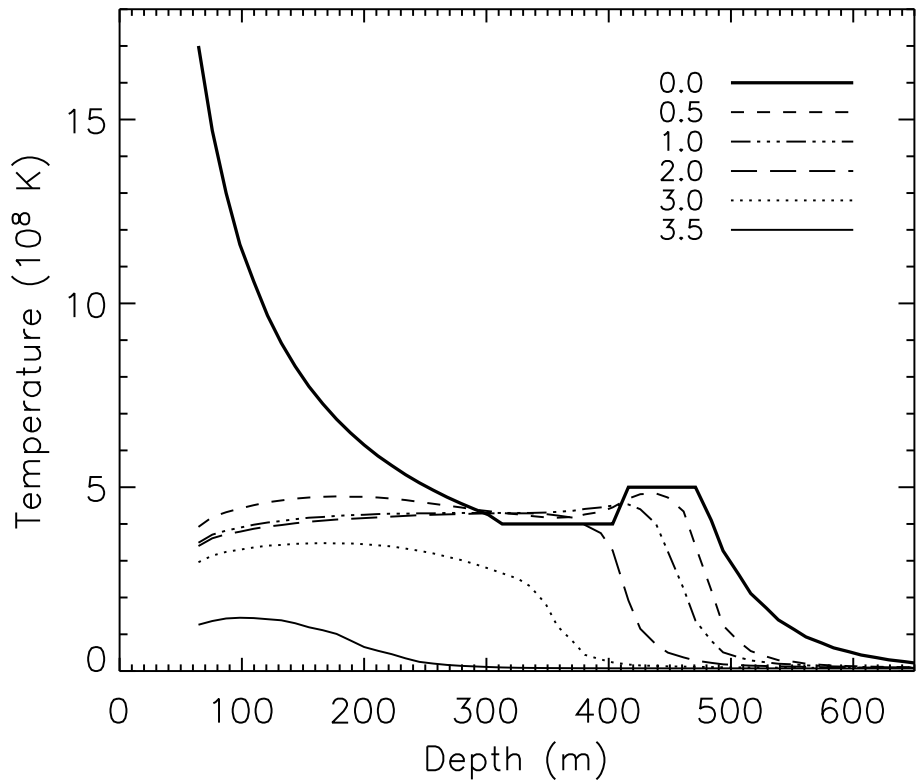}
%\vskip0.1truein

\caption{Evolution of temperature with depth, assuming a intial energy deposition of $10^{44}$ ergs. The different temperature profiles correspond to years after the initial injection, as indicated in the plot insert.}
\vspace{-0.5cm}
\end{figure}

The time behavior of the surface luminosity is displayed in Figure
1 (upper panel), where we have plotted the data points together with the theoretical
curve. It is seen that the plateau between days 400 and 800 is fit
very accurately. The reason is that the sharp rise in the specific
heat  near neutron drip and above $T_c$ makes this region in the
depth temperature plane a very large heat reservoir, which keeps
the temperature at neutron drip very stable. The duration of the
plateau is determined primarily by the time required for the inner
cooling wave to propagate outward to the neutron drip point. Most
of the heat in the inner crust is conducted to the center of the
star.  The surface lightcurve is insensitive to many other
details of the inner initial temperature profile, because by the
time the surface feels the latter, the spatial variations within
it are largely washed out.

This model is unable, of course, to explain the March 2003 data point, which showed that the flux did not decay further, and, moreover, showed a softening of  the spectrum. One could argue that the apparent flux ``ledge'' reflects a persistent, baseline luminosity  that is observed only when the star is sufficiently cool. Other SGRs in fact have larger persistent luminosities, which are temporarily buried by enhanced, transient afterglows that typically follow major bursting episodes. In the other cases, (such as with SGR~$1900+14$), the persistent luminosity was attributed (LET) to a hot core (temperature $\sim7 \times 10^8$ K). In the case of \sgr\, however, the ``observed'' (based, for the time being, on only two data points) persistent emission {\it cannot} be attributed to a hot core according to our model: we calculated that any core hot enough to account for observable persistent emission at the level of the last two points in Figure 2, would keep the bottom of the crust warm and thus smear out the observed sharp ledge in the light curve. Further, the surface temperature would not have dropped fast enough to have accounted for the first \chandra\ data point. However, if the persistent emission is from a small hot spot at the surface, it does not affect the cooling of the deep crust and can be considered superimposed onto the theoretical cooling curve. If this low level component is ever found to be pulsed, it would support our conjecture that the baseline emission is heating at the surface rather than at finite crustal depth. More closely spaced and longer observations are certainly needed to establish the current level and nature of the source emission.

\acknowledgements {We would like to thank the anonymous referee for an extremely rapid and insightful report, and Dr. M. H. Finger for many useful comments. C.K., P.W. and E.G. acknowledge support
from NASA grant NAG5-9350, \& SAO grant GO1-2066X. D.E. and Y.L.
gratefully acknowledge support from the Arnow Chair of Physics,
the Israel Basic Research Foundation, and the Israel-U.S.
Binational Science Foundation.}

\end{document}